\documentclass[preprint2]{aastex631}

\begin{document}

\title{The Red Supergiant Progenitor of the Type II Supernova 2024abfl}

\correspondingauthor{Zhengwei Liu, Xuefei Chen}
\email{zwliu@ynao.ac.cn, cxf@ynao.ac.cn}

\author{Jingxiao Luo}
\affiliation{Yunnan Observatories, Chinese Academy of Sciences (CAS), Kunming 650216, P.R. China}
\affiliation{University of Chinese Academy of Sciences, Beijing 100049, P.R. China}
\affiliation{International Centre of Supernovae, Yunnan Key Laboratory, Kunming 650216, P.R. China}

\author{Lifu Zhang}
\affiliation{Yunnan Observatories, Chinese Academy of Sciences (CAS), Kunming 650216, P.R. China}
\affiliation{University of Chinese Academy of Sciences, Beijing 100049, P.R. China}
\affiliation{International Centre of Supernovae, Yunnan Key Laboratory, Kunming 650216, P.R. China}

\author{Bing-Qiu Chen}
\affiliation{South-Western Institute for Astronomy Research, Yunnan University, Kunming, 650500, P.R. China}

\author{Qiyuan Cheng}
\affiliation{Yunnan Observatories, Chinese Academy of Sciences (CAS), Kunming 650216, P.R. China}
\affiliation{University of Chinese Academy of Sciences, Beijing 100049, P.R. China}
\affiliation{International Centre of Supernovae, Yunnan Key Laboratory, Kunming 650216, P.R. China}

\author{Boyang Guo}
\affiliation{Yunnan Observatories, Chinese Academy of Sciences (CAS), Kunming 650216, P.R. China}
\affiliation{University of Chinese Academy of Sciences, Beijing 100049, P.R. China}
\affiliation{International Centre of Supernovae, Yunnan Key Laboratory, Kunming 650216, P.R. China}

\author{Jiao Li}
\affiliation{Yunnan Observatories, Chinese Academy of Sciences (CAS), Kunming 650216, P.R. China}
\affiliation{International Centre of Supernovae, Yunnan Key Laboratory, Kunming 650216, P.R. China}

\author{Yanjun Guo}
\affiliation{Yunnan Observatories, Chinese Academy of Sciences (CAS), Kunming 650216, P.R. China}
\affiliation{International Centre of Supernovae, Yunnan Key Laboratory, Kunming 650216, P.R. China}

\author{Jianping Xiong}
\affiliation{Yunnan Observatories, Chinese Academy of Sciences (CAS), Kunming 650216, P.R. China}
\affiliation{International Centre of Supernovae, Yunnan Key Laboratory, Kunming 650216, P.R. China}

\author{Xiangcun Meng}
\affiliation{Yunnan Observatories, Chinese Academy of Sciences (CAS), Kunming 650216, P.R. China}
\affiliation{International Centre of Supernovae, Yunnan Key Laboratory, Kunming 650216, P.R. China}

\author{Xuefei Chen}
\affiliation{Yunnan Observatories, Chinese Academy of Sciences (CAS), Kunming 650216, P.R. China}
\affiliation{International Centre of Supernovae, Yunnan Key Laboratory, Kunming 650216, P.R. China}
\affiliation{Key Laboratory for the Structure and Evolution of Celestial Objects, CAS, Kunming 650216, P.R. China}

\author{Zhengwei Liu}
\affiliation{Yunnan Observatories, Chinese Academy of Sciences (CAS), Kunming 650216, P.R. China}
\affiliation{International Centre of Supernovae, Yunnan Key Laboratory, Kunming 650216, P.R. China}

\author{Zhanwen Han}
\affiliation{Yunnan Observatories, Chinese Academy of Sciences (CAS), Kunming 650216, P.R. China}
\affiliation{International Centre of Supernovae, Yunnan Key Laboratory, Kunming 650216, P.R. China}
\affiliation{Key Laboratory for the Structure and Evolution of Celestial Objects, CAS, Kunming 650216, P.R. China}



\begin{abstract}

Linkage between core-collapse supernovae (SNe) and their progenitors is not fully understood and ongoing effort of searching and identifying the progenitors is needed.
SN\,2024abfl is a recent Type II supernova exploded in the nearby star-bursting galaxy NGC\,2146, which is also the host galaxy of SN\,2018zd. 
From archival Hubble Space Telescope (HST) data, we have found a red source ($\mathrm{m_{F814W} \sim 25}$) near the location (angular distance $\leq 0.2"$) of SN\,2024abfl before its explosion.
With F814W and F606W photometry, we found that the properties of this source matched a typical red supergiant (RSG) moderately reddened by interstellar dust at the distance of the host galaxy.
We conclude that the SN\,2024abfl had an RSG progenitor with initial mass of $\mathrm{10M_{\odot}}$--$\mathrm{16\,M_{\odot}}$.

\end{abstract}



\section{Introduction} \label{sec:intro}

\begin{figure*}[ht!]
\includegraphics[width=1.6\columnwidth]{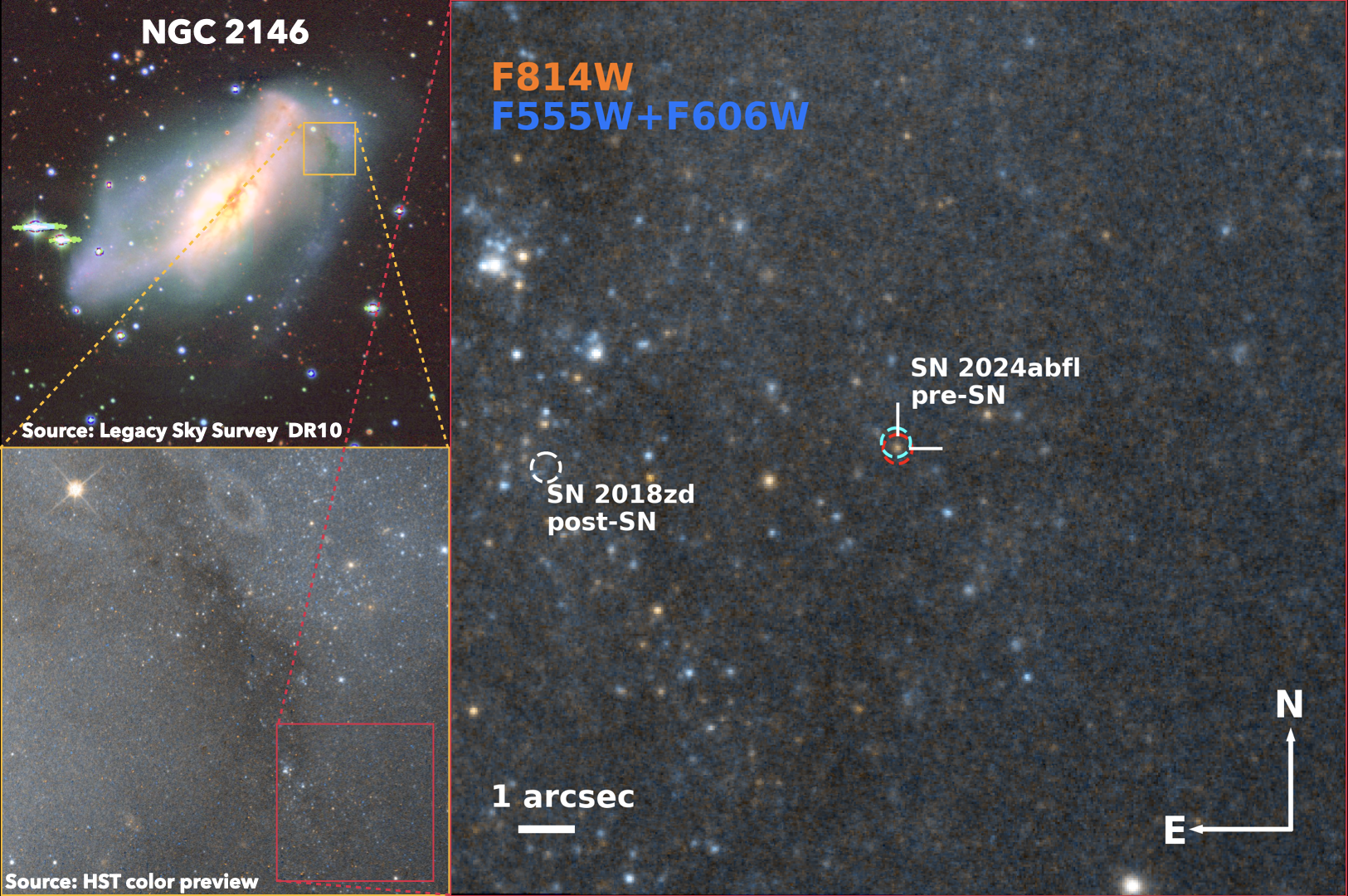}\centering
\caption{The zooming-in cutout of the SN\,2024abfl location. The upper left image is from Legacy Sky Survey DR10. The lower left image is a color preview of HST data which shows the spiral arm and dust lanes around SN\,2024abfl. On the right is the F555W+F606W/F814W color composite image centered on the location of SN\,2024abfl. The red and cyan dashed circles represent the location of SN\,2024abfl reported by ZTF and ATLAS team, respectively. The white dashed circle represents the location of SN\,2018zd reported by GaiaAlerts. The radius of the dashed circles are 0.25". The FOV of this cutout is 15.85" by 15.85". F555W and F606W are colored in blue, F814W is colored in orange.
\label{fig:color_combined}}
\end{figure*}

Type II supernovae (SNe II) are typically believed to be the outcomes of core collapsing massive stars with hydrogen-rich envelops \citep[see][]{2003ApJ...591..288H}.
They are crucial in physical processes such as galactic chemical evolution, star-formation feedback and formation of compact objects such as neutron stars and black holes. 
SNe II are also vital probes of stellar evolution theory as they reveals the structures of their massive progenitor at the end of their lives \citep[see][]{2013ARA&A..51..457N, 2012ARNPS..62..407J, 2012ARA&A..50..107L}.

In the era of Hubble Space Telescope (HST), finding progenitor stars for nearby SNe is possible on a regular basis \citep[see][for example]{2009MNRAS.395.1409S, 2010ApJ...714L.280F, 2011ApJ...742....6E, 2012ApJ...756..131V, 2014MNRAS.438..938M, 2018MNRAS.474.2116D} as HST allows high-resolution imaging of nearby galaxies, which is typically capable of resolving their supergiant populations.
However, the discoveries and identifications of SN progenitor stars are still limited as they required pre-explosion HST image at the SN location, which is not always guaranteed due to the small field of view (FOV) and overall footprint of HST.
When historical observations are available and deep enough to resolve the progenitors stars, links between SNe and progenitors stars can be made which has greatly improved our understanding of core-collapse supernovae (CCSNe) \citep[see][for reviews]{2009ARA&A..47...63S, 2017RSPTA.37560277V}, and have placed strict constrains on some Type Ia supernovae \citep[see][for example]{2011Natur.480..348L}.

\begin{table*}
	\centering
	\caption{Archival HST observations of SN\,2024abfl site, and photomtry results.}
	\label{tab:hst_logs}
	\begin{tabular}{ccccc} 
		\hline
		Observation Date & Filter & Exposure time & Proposal ID & Measured Magnitude \\
		(yyyy/mm/dd) &  & (s) & & \\
		\hline
		2004/04/10 & F814W & 120 & 9788 & $25.45\pm0.13$\\
		2004/04/10 & F658N & 700 & 9788 & \\
		2013/03/07 & F225W & 1500 & 13007 & \\
        2019/05/19 & F814W & 1320 & 15151 & $25.14\pm0.05$\\
        2019/05/19 & F555W & 1320 & 15151 & $27.48\pm0.21$\\
        2021/02/07 & F814W & 780 & 16179 & $25.26\pm0.07$\\
        2021/02/07 & F606W & 710 & 16179 & $27.18\pm0.23$\\
        2023/01/15 & F814W & 780 & 17070 & $24.89\pm0.04$\\
        2023/01/15 & F555W & 760 & 17070 & No Detection\\
        2024/01/31 & F555W & 1300 & 17506 & No Detection\\
        2024/01/31 & F275W & 1338 & 17506 & \\
		\hline
	\end{tabular}
\end{table*}

The identification of progenitor stars has played an important role in understanding massive star evolution and refined models of CCSNe.
During this process, more puzzles and peculiarities have surfaced recently.
One example is the Red Supergiant Problem, where the predicted progenitor upper mass limit for Type II-P supernovae (SNe II-P) differs from observations \citep[][]{2015PASA...32...16S}.
At the lower end of the SNe II-P progenitor mass range, electron-capture supernovae (ECSNe) may occur and it has been postulated that these kind of SNe happens when the core of super-Asymptotic Giant Branch (SAGB) stars undergoes electron capture process and promptly collapse into a neutron star \citep[][]{1980PASJ...32..303M, 1984ApJ...277..791N}.

The SN\,2018zd in the nearby galaxy NGC\,2146 was one of the first candidates for an ECSN \citep[][]{2021NatAs...5..903H}.
Its progenitor has been found in archival HST data which exhibited an abnormally low luminosity for a RSG, especially when combined with a closer-than-normal distance estimate for its host galaxy.
The post-explosion SN spectrum also showed features similar to that of an ECSN predicted by theory \citep[][]{2020MNRAS.498...84Z}.
Therefore, the progenitor of SN\,2018zd has been theorized as a SAGB which led to an ECSN.
However, the observation of the SN\,2018zd progenitor was based on only one detection in F814W filter, which forbid a more comprehensive analyses of the progenitor.

Coincidentally, six years later another SN II exploded just six arcsec away from SN\,2018zd.
SN\,2024abfl was first reported as a SN-candidate on November 15th, 2024 \citep{2024TNSTR4506....1I}.
The SN was quickly classified as a young and blue SN II on the same day by the HCT-2m telescope \citep{2024TNSCR4515....1D}.
On the following day a spectrum of SN\,2024abfl was obtained by Gemini-North with better resolution and signal-to-noise ratio (SNR) \citep{2024TNSCR4535....1A}. 
Thanks to the peculiarity of SN\,2018zd, this area of NGC\,2146 has been targeted by HST multiple times after 2018, and all these observations also includes the location of SN\,2024abfl, which enable us to study the progenitor of SN\,2024abfl in details.
From this archival HST dataset, we were able to identify the possible progenitor of SN\,2024abfl with four conspicuous detections in F814W band starting from 2004 (the same observation in which the progenitor of SN\,2018zd was found) to 2024, as well as low SNR detections in F606W and F555W bands.
These observations allowed us to better constrain the properties of the progenitor candidate, and can be used to place an upper-limit for pre-SN activities of the progenitor.

\begin{figure*}[ht!]
\includegraphics[width=2\columnwidth]{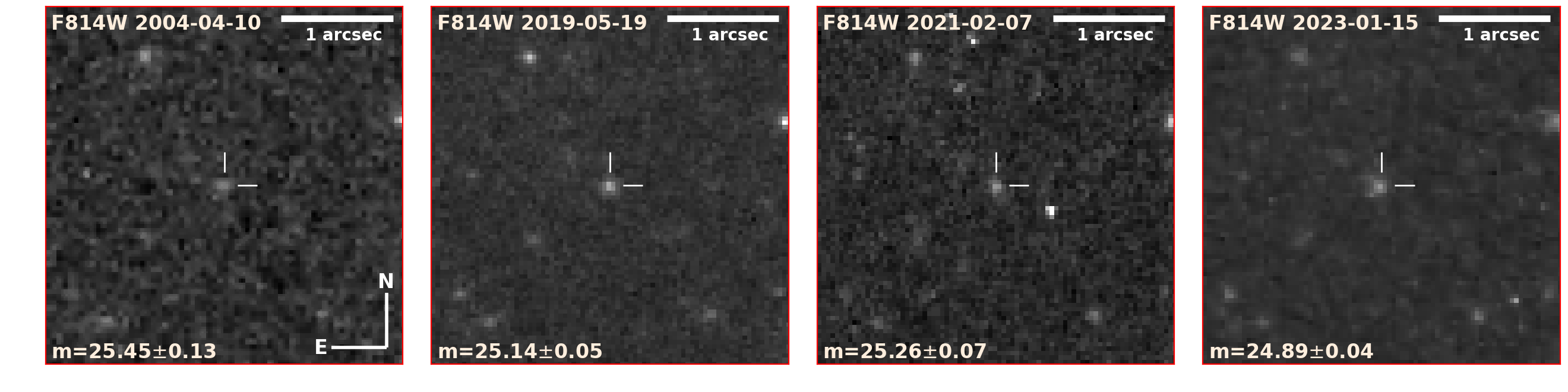}\centering

\includegraphics[width=2\columnwidth]{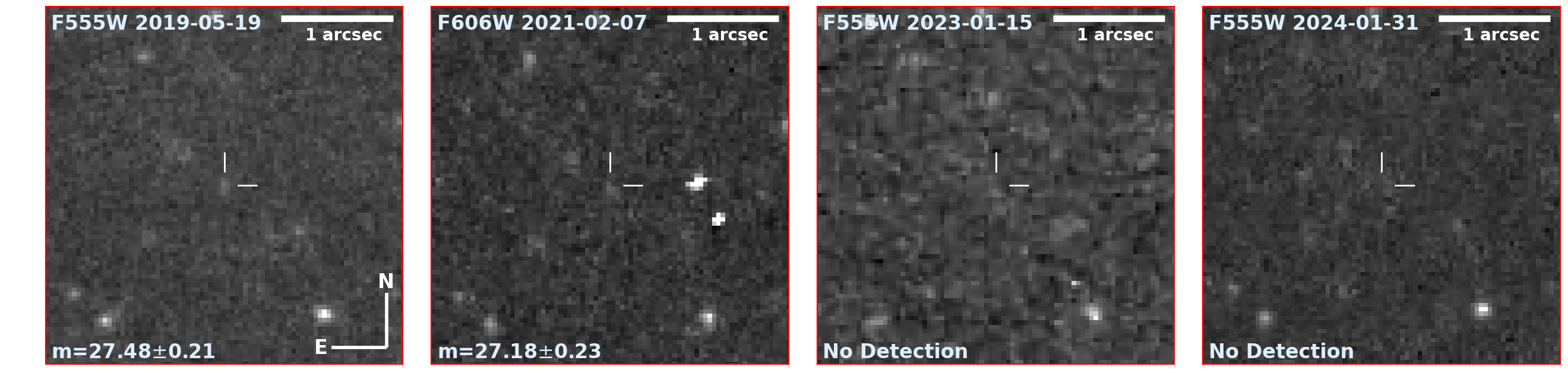}\centering
\caption{3.2" by 3.2" cutouts centered on SN\,2024abfl location. The cutouts on the upper row are from F814W observations. The cutouts on the lower row are from F555W and F606W observations.
\label{fig:Cutouts}}
\end{figure*}

The rest of this paper is organized as follows.
We first discuss details of past HST observations in Section~\ref{sec:Obs}. 
We then shows our method for identifying the progenitor candidate as an RSG and basic results in Section~\ref{sec:methods}.
We further discuss the properties of the progenitor candidate in Section~\ref{sec:diss}.
We conclude the paper in Section~\ref{sec:conc}.

\section{HST Observations} \label{sec:Obs}

All of the data used for our analyses was downloaded from Mikulski Archive for Space Telescopes (MAST)\footnote{\url{http://mast.stsci.edu/}}. There were about a dozen of HST observations focused on NGC\,2146 and six of them captured the location of SN\,2024abfl (with 13 individual exposures from ACS and WFC3 with various filters). The observation date, filters used, exposure times and other details including our magnitude measurement results are summarized in Table~\ref{tab:hst_logs}.

We have combined all F555W, F606W and F814W observations (except for the F814W observation from 2004 due to its relative short exposure time) into a composite color image centered on the location of SN\,2024abfl. 
The FOV of this colored image is 15.85" by 15.85", which also includes the location of SN\,2018zd.
This cutout and it surrounding field showing the entire NGC\,2146 galaxy (image from Legacy Sky Survey\footnote{\url{https://www.legacysurvey.org/viewer/}}) is shown in Fig.~\ref{fig:color_combined}.

\footnote{All the {\it HST} data used in this paper can be found in MAST: \dataset[10.17909/hp6p-0735]{http://dx.doi.org/10.17909/hp6p-0735}.}

\section{Methods} \label{sec:methods}

\subsection{Photometry and Results} \label{subsec:photometry}

We did not redo the reduction and calibration processes for the HST data and used the calibrated science-level data available on MAST as is. Fig.~\ref{fig:color_combined} shows a red stellar object at the location of SN\,2024abfl reported by ground-based surveys.
The centroid of this object aligns well with the location of SN, with error less than 0.2 arcsec.
We assume this stellar object as the likely progenitor candidate for SN\,2024abfl and used the python package ASTROPY and its photometry module photutils to measure magnitudes of this progenitor candidate.
The aperture radius we used for flux measurement was 0.15 arcsec, together with inner and outer annulus radius of 0.4 and 0.6 arcsec for background subtractions.
This setting is also used for detecting other nearby sources in our attempt of generating a color-magnitude diagram (CMD) for stellar population of NGC 2146.
A chance alignment of this object with SN\,2024abfl is possible, but unlikely and requires complex setups to fit the observed SN.
We shall discuss other possibilities in Section~\ref{sec:diss}.

The progenitor candidate was clearly visible in all F814W images, while in shorter wavelengths (F555W and F606W) it was barely detectable, and it was well below detection limit in F225W, F275W and F658N bands.
We summarize the photometry results in Table.~\ref{tab:hst_logs}, and 3.2 arcsec by 3.2 arcsec cutouts centered on the SN\,2024abfl location in Fig.~\ref{fig:Cutouts}.
From the six photometric measurements available, we can plot the pre-explosion light curve of the progenitor candidate in Fig.~\ref{fig:lc}.
Apart from a slight increase in F814W flux in 2023, the overall appearance of the progenitor candidate was stable and there was no sign of dramatic brightening or dimming in these wavelengths.

\subsection{Evolution Tracks} \label{subsec:evo}

We used MESA version r22.05.1 \citep[][]{2011ApJS..192....3P, 2013ApJS..208....4P, 2018ApJS..234...34P} to generate evolution tracks of RSG with initial mass ranging from $\mathrm{10\,M\odot}$ to $\mathrm{25\,M\odot}$, with initial metallicity of Z\,=\,0.02.
Other settings include overshooting coefficient $\mathrm{f_{ov}\,=\,0.014}$, $\mathrm{\alpha_{mlt}\,=\,1.8}$, Dutch wind \citep[][]{1988A&AS...72..259D, 2001A&A...369..574V} factor\,=\,0.8 and nuclear reaction network 'sagb\_NeNa\_MgAl.net'.
We terminated our MESA runs at the time of carbon depletion in the core, and use the stellar models at that time to compare with our observations.

\subsection{Distance Estimates} \label{subsec:distance}

We noticed that the exact distance of the host galaxy NGC\,2146 has been a debated topic on some post-explosion analysis of SN\,2018zd \citep[see][for example]{2021arXiv210912943C}.
NGC\,2146 is unfortunately located at a distance where its apparent redshift (i.e. receding velocity) is not accurate enough due to possible peculiar velocity of this galaxy affecting the redshift-distance relation as it is too close.
While in the meantime NGC\,2146 was also too far-away for accurate Cepheid or TRGB based distance measurement is available.
We decide to adopt a distance estimate of $\mathrm{15.6^{+6.1}_{-3.0}\,Mpc}$ or $\mathrm{D\,=\,12.6\,-21.7\,Mpc}$ according to analysis in \cite{2021arXiv210912943C}, and discuss our result based on the lower and upper end of this estimate in the following sections.
Its out of scope of this study but we hope future studies can derive a better distance estimate for NGC\,2146.

\begin{figure}[ht!]
\includegraphics[width=1.1\columnwidth]{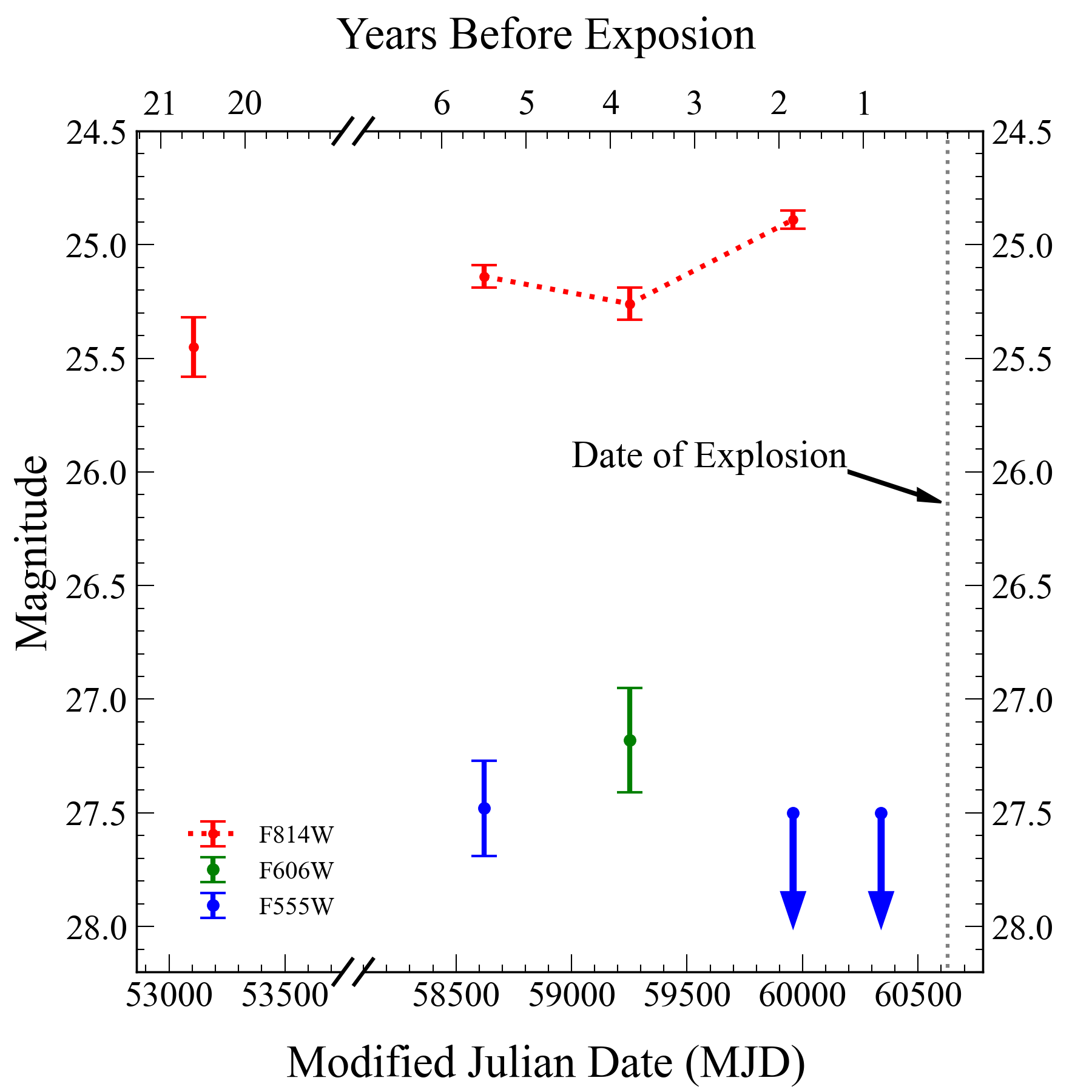}\centering
\caption{Light curve of SN\,2024abfl progenitor candidate.
\label{fig:lc}}
\end{figure}

\section{Discussion} \label{sec:diss}

\subsection{A reddened RSG} \label{subsec:rsg}
\begin{figure}[ht!]
\includegraphics[width=0.9\columnwidth]{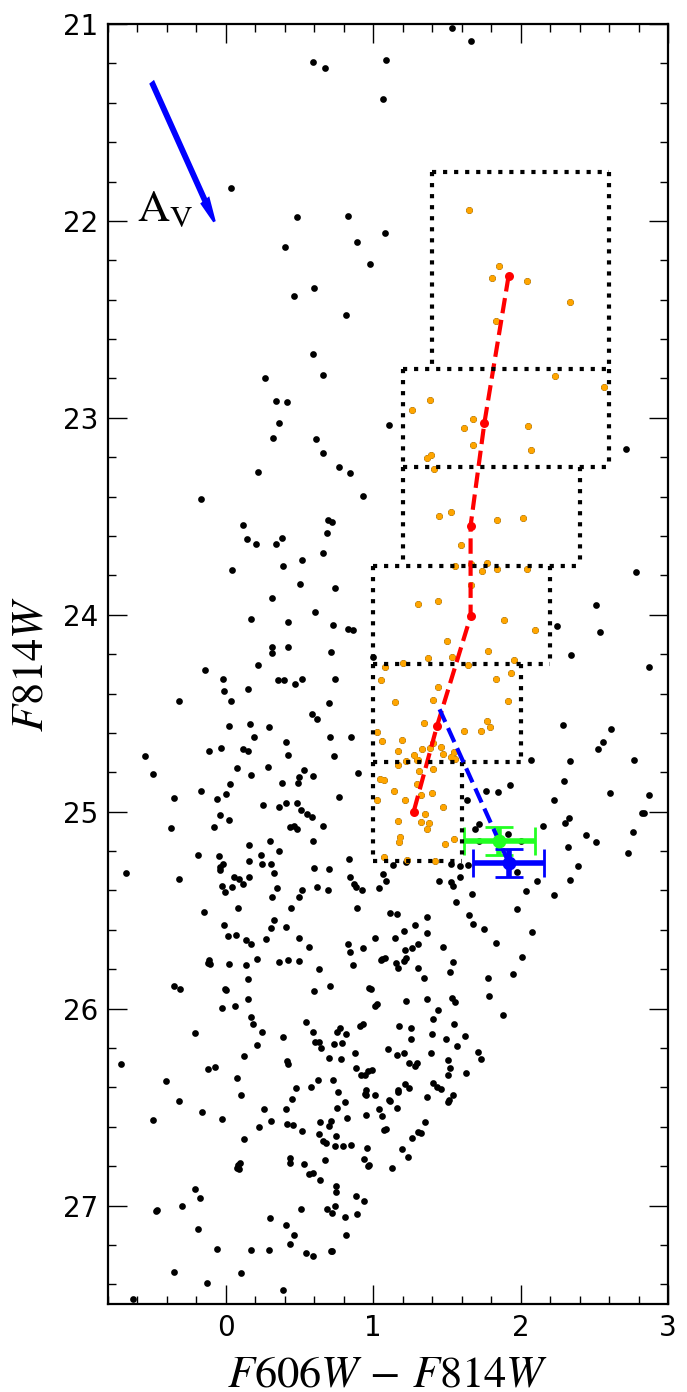}
\caption{Progenitor of SN\,2024abfl on the color-magnitude diagram. The background dots are nearby stars in its host galaxy. The position of the progenitor is shown as the blue dot with error bars. If shifted back along the direction of extinction vector $\mathrm{A_{F814W, host}\,\sim\,0.1}$ indicated by the NaID absorption, then the progenitor would be at the green dot. The red dashed line represent the center line of the RSG branch in the host galaxy. The orange dots are stars used to calculate the position of RSG branch. The blue dashed line shows the direction of movement if one tries to shift the progenitor back to the RSG branch along the anti-direction of extinction vector.
\label{fig:cmdiagram}}
\end{figure}

Before applying any extinction correction, an apparent F814W magnitude of about 25.2 translate to $\mathrm{M_{F814W}\sim-5.3}$ at a distance of $\mathrm{D\,=\,12.6\,Mpc}$, or $\mathrm{M_{F814W}\sim-6.5}$ at $\mathrm{D\,=\,21.7\,Mpc}$, which roughly correspond to the expected brightness of a dim, less massive RSG.
Apart from F814W magnitudes, we also detected this source in shorter wavelength, allowing us to plot it in a CMD. 
In Fig.~\ref{fig:cmdiagram} we plot the progenitor candidate against a background CMD of NGC\,2146 extracted from archival HST observations in F814W and F606W.
There are 576 star-like sources with both F606W and F814W detections in this CMD, and most of the them are located on or near the disk and spiral-arm of NGC 2146.
On the CMD two branches of stars are visible, the left branch centered on F606W-F814W $\sim$ 0.5 consisted of massive main sequence stars, while the branch on the right is the RSG branch.
More details on Fig.~\ref{fig:cmdiagram} is described in the Appendix.

The progenitor candidate lies slightly below and to the right of the RSG branch.
If we assume the progenitor candidate was a RSG, then its location on the CMD indicates that it was moderately reddened, with a total $\mathrm{A_{F814W}\,\sim\,0.8}$.
The Gemini spectrum of SN\,2024abfl post explosion (available on TNS) also supports an additional reddening, in which two sets of narrow NaID absorption lines can be found, one centered on zero redshift and the other centered on the redshift of NGC\,2146.
We tried to fit those two lines with Gaussian profiles and calculated their equivalent widths to be 0.77 \r{A} (Milky Way) and 0.58 \r{A} (host galaxy).
According to \cite{2012MNRAS.426.1465P}, this amount of extinction roughly correspond to $\mathrm{A_{F814W, host}\,\sim\,0.1}$, which is lower than what we inferred from the CMD.
However, the correlation between NaID equivalent widths and actual amount of extinction has a significant scatter \citep[see][]{2011MNRAS.415L..81P}.
Therefore, it only serves as a constrain to show that the extinction suffered by SN\,2024abfl was moderate, and blue supergiant progenitor is unlikely as it requires heavy amount of extinction to match the pre-SN HST observations.

\begin{figure*}[ht!]
\includegraphics[width=1.85\columnwidth]{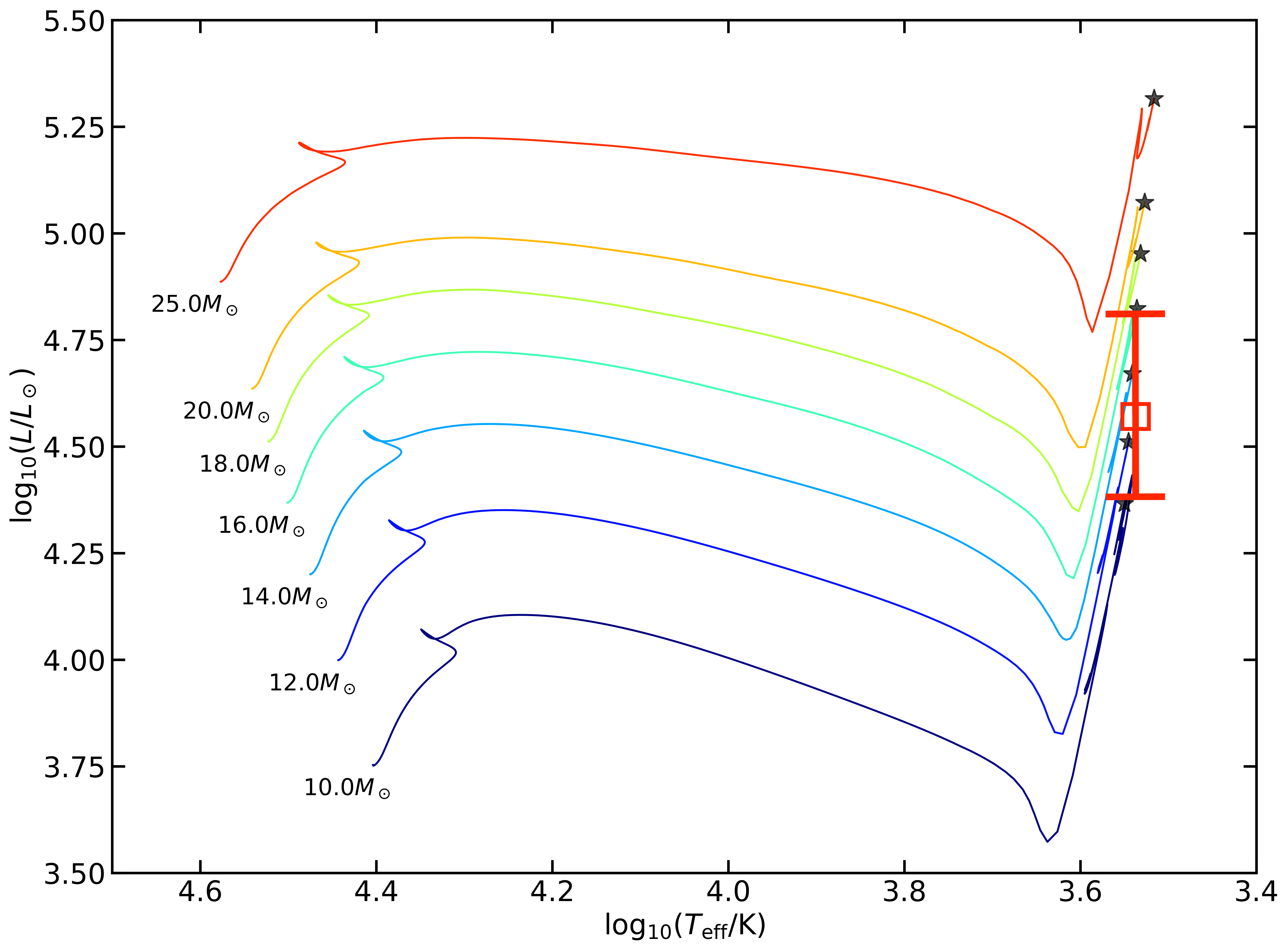}\centering
\caption{MESA evolution tracks of massive star with initial mass range of $\mathrm{10M_{\odot}}$ to $\mathrm{25M_{\odot}}$. The red box shows the luminosity of SN\,2024abfl progenitor at a distance of $\mathrm{D\,=\,15.6\,Mpc}$, the error bar shows the luminosity range corresponding to $\mathrm{15.6^{+6.1}_{-3.0}\,Mpc}$. The gray marks are the positions when carbon in the core is exhausted.
\label{fig:MESA}}
\end{figure*}

In order to understand whether the additional reddening came from circum-stellar material (i.e. similar to SN\,2023ixf \citep[][]{2023ApJ...957...64S, 2024ApJ...968...27V,2023ApJ...955L..15N}) or interstellar dust, we searched for infrared excess in archival SPITZER data and found no infrared counter-part at the location of the progenitor.
This suggests that the majority of reddening should come from environmental dust within the spiral arm where SN\,2024abfl exploded.
Overall, the properties of the progenitor can be explained by a reddened RSG on the lower part of RSG SN progenitor mass range.
To get the bolometric luminosity of the progenitor in order to compare it with evolution tracks, an additional step of bolometric correction is required.
We adopt a suggested BC=0.0 by \cite{2024arXiv241014027B} for the progenitor as it is a RSG approaching core-collapse, which may be closer to a spectral type M5 giant than an M0 giant.

When adopting BC=0.0 and corrected for the estimated total extinction, the F814W brightness of the progenitor translate to a luminosity of $\mathrm{logL\,\sim\,4.32}$ at $\mathrm{D\,=\,12.6\,Mpc}$, or $\mathrm{logL\,\sim\,4.8}$ at $\mathrm{D\,=\,21.7\,Mpc}$.
Therefore, we can plot the luminosity of the progenitor along with evolutionary tracks of massive stars with initial mass range of $\mathrm{10M_{\odot}}$ to $\mathrm{25M_{\odot}}$ in Fig.~\ref{fig:MESA}.
The luminosity estimates of the progenitor star roughly correspond to the luminosity range of $\mathrm{10M_{\odot}}$ to $\mathrm{16M_{\odot}}$ stars when they exhaust their carbon in the core.

\subsection{Brightening and Outburst} \label{subsec:outburst}

As displayed in Fig.~\ref{fig:lc} and Table.~\ref{tab:hst_logs}, the F814W magnitudes of the progenitor before SN explosion shows a general upward trend.
In F555W band, the source seemed to have faded out of HST detection, at least it has not brightened significantly in the last two F555W observations.
These behaviors indicate a potential subtle brightening of the progenitor years before the final explosion.
At the same time, major pre-SN outburst of the progenitor during or shortly before the HST observation epochs can be ruled out.

Comparing the 2019 and 2023 observations in F814W and F555W, the F814W flux of the progenitor was increased by about 25 percent.
We tried to estimate the detection limit of the 2023 F555W observation by doing forced photometry at the location of the progenitor, which turned out to be $\mathrm{26.9\pm0.2}$ despite no significant source at that location was visible.
If we assume the color of progenitor remained unchanged (i.e. no significant cooling), then its F555W magnitude in 2023 would have been 27.2, still below the detection limit.
This means that we cannot confirm any rapid expansion and consequently cooling of the progenitor's envelop a few years before explosion, as the sensitivity of the observation in F555W was simply not enough.

We note that slight fluctuations in brightness are not uncommon among RSGs (for example, Betelgeuse).
It is possible that the progenitor had some kind of pulsation, and HST happened to capture the star at a particular phases to show an upward trend in these observations.
Whether this slight increase in luminosity (and potential drop in temperature in 2023 and 2024) is a result of terminal-stage nuclear burning or pulsations of the envelop remains unclear.

\subsection{Other Possibilities and Future Observations} \label{subsec:others}

We notice that in order to fit the current observation with a RSG according to Section.~\ref{subsec:rsg}, an additional reddening of $\mathrm{E(F606W-F814W)\,\sim\,0.4}$ is required.
This means that the host extinction might be a few times stronger than galactic extinction, which is not evident in post-SN spectra that shows almost equal width of galactic and host NaID doublet.
If we adopt the host extinction from the NaID equivalent width from \ref{subsec:rsg}, then the F606W-F814W color of the progenitor will no longer fit that of a RSG.
Instead, the progenitor might be an luminous SAGB, which exploded as an ECSN in this case.
This possibility might be confirmed or ruled out by future spectroscopy and photometry observation of SN\,2024abfl, which will show whether this SN exhibits features predicted for an ECSN.

It is not likely for the progenitor to be a heavily-reddened blue supergiant as it would require far greater amount of extinction to shift a blue spectrum into one with red F606W-F814W color.
At the same time, the early spectra of SN\,2024abfl are blue, making the heavily-redden scenario less plausible.

It is also possible that the SN exploded in a binary system, which means that the progenitor we detect consists of flux from the actual SN progenitor and an additional, non-related source.
Similarly, a chance alignment of the actual progenitor and a background or foreground star within the host galaxy is also possible.
Eventually another high-resolution, deep observation of the SN site by HST or JWST after the SN has faded will be required to rule out these possibilities.
These observations will help confirm either the disappearance of the progenitor star, or a significant decrease in flux which indicates a member of a binary system has exploded.

In addition, as SN\,2018zd was just six arcsec away from SN\,2024abfl, any future HST or JWST observations will be able to observe the location of SN\,2018zd simultaneously, which may provide additional information about SN\,2018zd (e.g. neighboring stars, early supernova remnant, etc.).
Deep observation of NGC\,2146 in general is also important as it may provide possibility of a more accurate distance measurement bases on techniques such as TRGB or Cepheid period-luminosity relation.

As of now, the SN\,2024abfl itself is also exhibiting uncommon behaviors, as the brightness of the SN seems to have never reached mag 16.5 and above.
SN\,2024abfl maintained brightness of mag ~17 since explosion and a peak in brightness seems to be absent.
This brightness is lower than the expected brightness of SNe II, even when the reddening estimated in the previous sections being accounted for.
Therefore, SN\,2024abfl might be a rather peculiar SN in itself, which means that more observations from ground-based telescopes might be equally interesting and scientifically important.

\section{Conclusions} \label{sec:conc}

From archival HST observations, we have found a star-like object at the location of SN\,2024abfl which exploded recently.
The source had an average magnitude in F814W band of $\mathrm{m_{F814W}\,\sim\,25.2}$, and in F606W/F555W bands it was just above detection limit with a magnitude of around 27.
The color and brightness of this source can be explained by a moderately reddened RSG at the distance of SN\,2024abfl host galaxy NGC\,2146.
We therefore identify this source as a candidate for progenitor of SN\,2024abfl, with a estimated mass of $\mathrm{10M_{\odot}\,\sim\,16M_{\odot}}$.
Future observations such as spectrum at plateau and nebular phase of SN\,2024abfl will reveal more information on the SN itself.
Currently the biggest part of error came from the large scatter in the host galaxy distance measurement, we hope deep HST or JWST observation in the future will help determine the distance of the host galaxy via Cepheid variables or TRGB measurements.

\begin{acknowledgments}
This work is supported by the National Natural Science Foundation of China (NSFC, Nos. 12288102, 12125303 , 12090040/1, 11873016), the National Key R\&D Program of China (Nos. 2021YFA1600401 and 2021YFA1600400), the Chinese Academy of Sciences (CAS), the International Centre of Supernovae, Yunnan Key Laboratory (No. 202302AN360001), the Yunnan Fundamental Research Projects (grant Nos. 202201BC070003, 202001AW070007) and the “Yunnan Revitalization Talent Support Program”–Science and Technology Champion Project (No. 202305AB350003).

\end{acknowledgments}

%

\vspace{5mm}
\facilities{HST(ACS), Swift(XRT and UVOT), Spitzer}


\software{astropy \citep{2013A&A...558A..33A,2018AJ....156..123A},
          photutils \citep{larry_bradley_2024_12585239}
}



\appendix

\section{Comparing with known RSG spectra}

\begin{figure}[ht!]
\includegraphics[width=0.7\columnwidth]{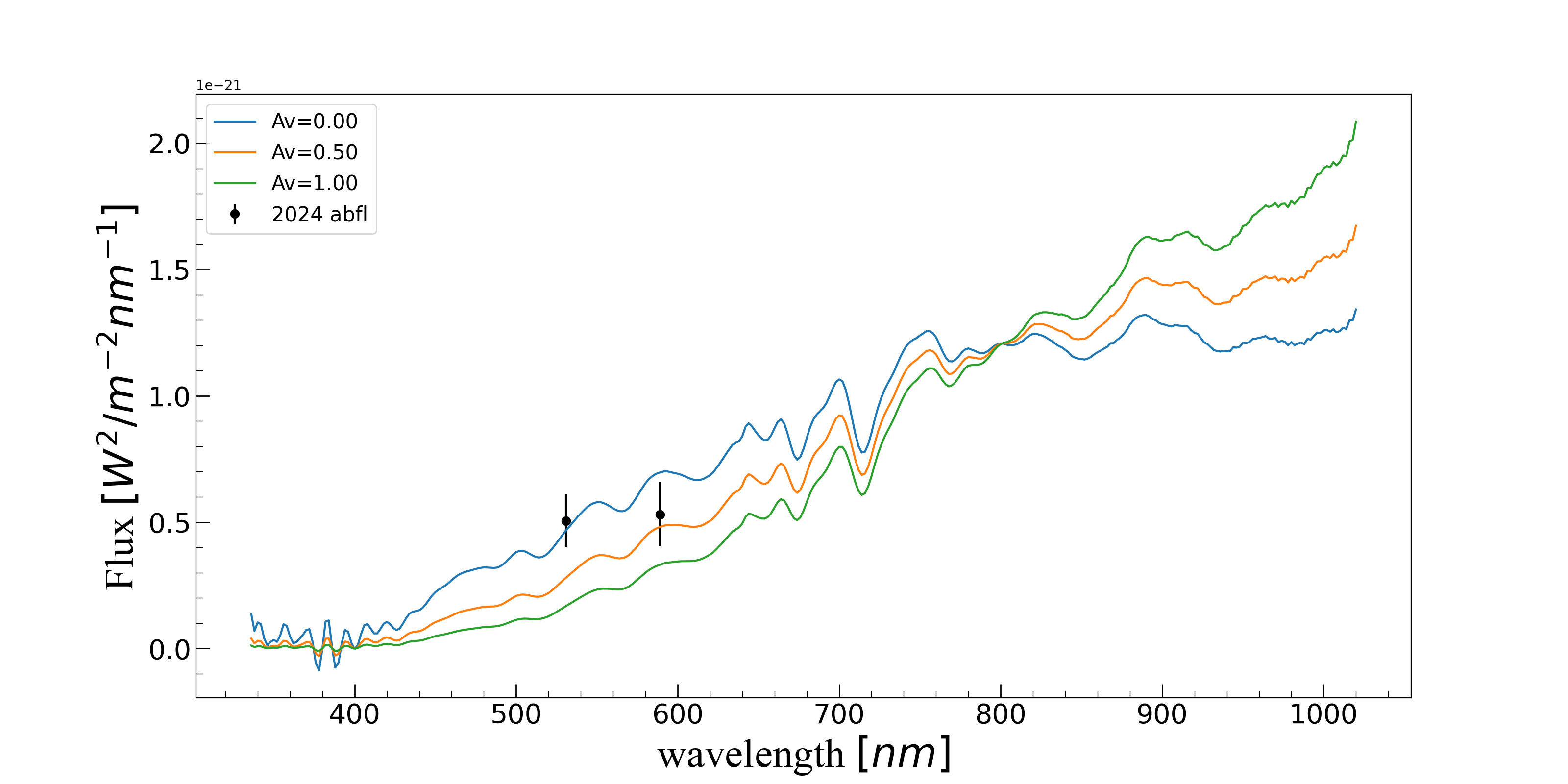}\centering
\caption{Spectra of LMC002 with different amount of extinction, normalized to have the same F814W as the SN\,2024abfl progenitor. The F606W and F555W magnitudes of the progenitor is also marked. The central wavelength of each filter is placed at the pivot wavelength given by the HST FITS header.
\label{fig:spec}}
\end{figure}

To better illustrate that the progenitor of SN\,2024abfl is indeed an moderately reddened RSG, we compared our multi-color photometry result with a spectrum of RSG in LMC \citep[][]{2015A&A...578A...3G} in Fig.~\ref{fig:spec}.
From this comparison, it is obvious that a non-zero reddening is required to explain the color of the progenitor.
We note that the spectral type of this LMC RSG is M3, and is likely a high-mass, luminous RSG, which may be slightly different from the low-mass, less-luminous progenitor of SN\,2024abfl.

\section{Color-Magnitude Diagram of NGC 2146}

The CMD of NGC\,2146 was generated from archival HST data (Proposal ID: 16179) observed on Feb.7th, 2021, with F814W and F606W filter.
West part of NGC\,2146 covered half of FOV, and the other half of FOV provided a good sample of galactic foreground and background.
Preliminary source detection was preformed and a technique similar to the HST CI-index was used to remove hot pixels and extended sources (e.g., background galaxies or clusters within NGC\,2146).
Apart from the 0.15 arcsec aperture mentioned before, we also used a 0.05 arcsec aperture.
The flux ratios between these apertures can then be computed to identify real stellar objects.
Only sources with $0.65\,<\,m_{0.05"}\,-\,m_{0.15"}\,<\,1.25$ were used to generate the CMD.
Also we excluded all sources with $m\,>\,28.5$, leaving those that have detectable fluxes in both bands.
We then separated detected stars on the RSG branch into six bins according to their F814W magnitudes, and compute the centroid of each bin to generate the spline of the RSG branch in order to quantitatively define its location.


\bibliography{sample631}{}
\bibliographystyle{aasjournal}



\end{document}